\documentstyle[preprint,aps]{revtex}
\begin{document}
\title { Is the $\eta $ Meson a Goldstone Boson? }
\author{M. KIRCHBACH}
\address{ Institut f\"ur Kernphysik, Johannes Gutenberg Universit\"at,
D--55099 Mainz, Germany}
\author{H.J. WEBER}
\address{ J. W. Beams Laboratory and Department of Physics, University
of Virginia, \\Charlottesville, VA 22901, USA}
\maketitle
\begin{abstract}
The decoupling of the $\eta $ meson from the nucleon, as recently
deduced from analyzing $\bar p p$ collisions and $\eta $
photoproduction off the proton at threshold, is shown to provide an
argument against the octet Goldstone boson nature of the
$\eta $ meson. This argument concerns the structure of the strong
isoscalar axial vector current. A vanishing $\eta N$ coupling
means a vanishing contribution of the $\eta $ pole term to the
hypercharge nucleon axial vector current. Therefore,
no partial conservation of the latter can be achieved.
This new situation invalidates the octet Goldberger--Treiman
relation for the $\eta N$ coupling constant that is indicative of
the octet Goldstone boson nature of the $\eta $ meson.
In such a case there is no longer a compelling reason
for the standard belief that the $\eta $ meson should be coupled to
the hadronic vacuum through the hypercharge axial vector current.
Rather, it will be coupled to it as any ordinary neutral pseudoscalar 
meson via the neutral axial vector current of the electroweak theory.
As a result of the suggested universal structure of the weak and strong 
neutral axial vector currents the $\eta $ meson satisfies the strange 
analog of the Goldberger-Treiman relation so that it acquires features 
of a strange Goldstone boson. Hence the $\eta NN $ vertex constant 
appears proportional to $\Delta s$, the fraction of nucleon spin carried 
by the strange quark sea, thus explaining its suppression naturally.

\end{abstract}

\section{Standard View Of $\pi^0$ And $\eta$ Mesons }
\par
In the limit of zero masses of the up (u) and down (d) quarks,
quantum chromodynamics (QCD) is invariant under global chiral rotations
of the quark field,
$q\rightarrow \mbox{exp}(-{i\over 2} \vec \alpha \cdot \vec\tau \gamma_5) q$,
$\bar q\rightarrow \bar q \mbox{exp}(-{i\over 2}\vec \alpha \cdot \vec\tau
\gamma_5)$. When chiral symmetry is broken spontaneously Goldstone bosons
arise that can be exploited as effective degrees of freedom.
\par
The simplest example for a chiral field theory is the Gell-Mann--L\'evy
$\sigma$-model\cite{GML} with a Mexican hat-type potential for a (fictitious)
scalar $\sigma$-meson and the pions. Its infinitely degenerate ground states
lie in the rim of the hat and are marked by a corresponding
non-zero vacuum expectation value
$<\sigma>_0 $ of the $\sigma$-meson. Massless current
quarks become massive (constituent) with  $m_q\sim <\sigma>_0$
in the broken symmetry phase when one of the vacua is spontaneously
selected. The rim is flat in the pion directions which
means that pions still have zero mass. They are the corresponding Goldstone
bosons \cite{GSW}. More realistic descriptions of chiral hadron systems
are provided by Nambu--Jona-Lasinio models that are patterned after
superconductivity \cite{NJL}. The effective degrees of
freedom are Goldstone bosons and dynamical quarks whose momentum dependent
mass $m_q(p^2)$ originates from a gap equation. Upon approximating the
dynamical mass by $m_q(0)\approx m_N/3$ one can introduce the concept of a
constituent quark of the nonrelativistic quark model (NQM) at low momentum.
\par
The Noether currents of the two--flavor chiral group
$SU(2)_L\otimes SU(2)_R$ coincide with the isovector axial vector currents 
of the Glashow--Weinberg--Salam (GWS) electroweak gauge theory
\begin{eqnarray}
J_{\mu , 5}^{(i)} &=&
\bar q  \gamma_\mu \gamma_5  {\lambda^i\over 2}q\, ,i=1,2,3.
\label{trplt_axcr}
\end{eqnarray}
They couple the corresponding Goldstone bosons, the pions, to the hadronic
vacuum in a Lorentz invariant way, e.g. for the neutral pion
\begin{equation}
\langle 0|\bar q \gamma_\mu \gamma_5{\lambda^3\over 2} q |\pi^0
\rangle = f_{\pi^0} m_{\pi^0} \, ik_\mu \, ,
\label{f_pi}
\end{equation}
where $k_\mu $ is the pion four--momentum, and
$f_{\pi^0}m_{\pi^0}=85\pm0.3$ MeV is the {\it ordinary } weak
decay constant of the neutral pion \cite{PART}, while $f_{\pi^0}$
stands for the respective {\it dimensionless } decay constant.
Such non--vanishing matrix elements are signals of spontaneous chiral
symmetry breakdown. They also enter into the weak decays of pions, e.g.
the $\pi^+ \to \mu^+ + \nu _{\mu}$ decay as well as the $\pi^0$
decay into two photons through the neutral axial vector current at one
corner of an intermediate quark triangle diagram with photons at the 
other two vertices, a mechanism known as chiral anomaly \cite{ABJ}.
It shows up in the divergence of the neutral axial vector current for
massless quarks as
\begin{equation}
 \partial^\mu J^{(3)}_{\mu ,5} = {n_3\alpha\over 2\pi}
             F_{\mu \nu}\tilde F^{\mu \nu}\, ,
\label{chiranom}
\end{equation}
where
$\tilde F^{\mu\nu}={1\over 2}\epsilon^{\mu\nu\rho\sigma} F_{\rho\sigma}$
is the dual of the electromagnetic field tensor, and
$n_3=Tr(\tau^3 Q^2)=3[({2\over 3})^2-(-{1\over 3})^2]=1$
is the charge factor corresponding to u and d triangles including colors.
\par
The generalization to $SU(3)_L\otimes SU(3)_R$ chiral rotations in
three flavor quark space (u,d,s) as symmetry transformations of the QCD
Lagrangian describing low mass hadron systems is one of the basic 
paradigms of contemporary hadron physics that has its roots partly in the 
success of the quark model with the SU(3) flavor octets. There are then 
three different neutral axial vector Noether currents corresponding to 
chiral rotations \cite{Col}. One additional Noether
current is the hypercharge axial vector current, $J_{\mu ,5}^{(8)}$, that
contains $\lambda^8$. It is expected to couple to the $\eta$ meson similar
to Eq.~(\ref{f_pi}) thereby defining the dimensionless $\eta$ decay 
coupling constant $f_\eta$.
\par
In contrast, the Noether current associated with the $U_A(1)$ symmetry
of the QCD Lagrangian is the singlet axial vector current with
$\lambda^0=\sqrt{{2\over 3}}\hat 1$ in Eq.~(\ref{trplt_axcr}) instead of
$\lambda^i$. Its divergence is given by
\begin{equation}
 \partial^\mu J^{(0)}_{\mu ,5} = {N_f \alpha_s\over 4\pi}
             F^{c}_{\mu \nu}\tilde F_{c}^{\mu \nu},
\label{anom}
\end{equation}
which has the $U_A(1)$ anomaly containing the gluon field strength
$F^c_{\mu \nu}$ and its dual $\tilde F_c^{\mu \nu}$.
In spite of the spontaneous breakdown of the $U_L(1)\otimes U_R(1)$ 
symmetry, it was suggested by 't Hooft that no corresponding Goldstone 
boson arises because of instanton effects. Thus, the properties of the 
$\eta'$ meson differ substantially from those of Goldstone bosons, and 
it is not considered here further.
\par
In addition, a purely strange axial vector current $J_{\mu ,5}^{(s)}$ 
can be defined through the ideal mixing between the hypercharge and
singlet axial vector currents as
\begin{equation}
J_{\mu , 5}^{(s)} =
-{1\over \sqrt{2}}\, (\sqrt{{2\over 3}} J^{(8)}_{\mu ,5}
-\sqrt{{1\over 3}}J^{(0)}_{\mu ,5})
= {1\over 2} \bar s \gamma_\mu \gamma_5 s\, .
\label{str_curr}
\end{equation}
\par
The matrix elements of the, say, isovector axial current
for the nucleon state is parametrized as \cite{KaMa}:
\begin{equation}
\langle N|\bar q\gamma_\mu \gamma_5 {\lambda^3\over 2} q|N\rangle
= \bar {\cal U}_N
[g_A(q^2)\gamma_\mu +q_\mu G_2(q^2) ] \gamma_5 {\tau^3\over 2}
{\cal U}_N \, , \quad \quad  g_A (0) = \Delta u -\Delta d\, ,
\label{gA}
\end{equation}
where $g_A (q^2) $ denotes the weak isovector axial form factor of 
the nucleon. Upon replacing $\lambda^3$, $g_A$, and $G_2$ in
Eq.~(\ref{gA}) by $\lambda^{i}$, $g_A^{(i)}$, and $G_2^{(i)}$ with
$i= 8,0$, respectively, one obtains
similar equations for the hypercharge and singlet axial form factors
involving the hypercharge and singlet axial coupling constants
\begin{equation}
g_A^{(8)} (0)  =  {1\over \sqrt{3}} (\Delta u +\Delta d
-2\Delta s)\, ,\qquad
 g_A^{(0)}(0)  = \sqrt{{2\over 3}} (\Delta u +\Delta d + \Delta s)\, .
\label{gA_0}
\end{equation}
Here $\Delta u$, $\Delta d $, and $\Delta s$ denote the fraction of
proton spin carried by the $u$, $d$, and $s$ quarks, respectively.
Just for completeness, we introduce the nucleon matrix
element of the purely strange axial vector current as
\begin{equation}
\langle N| \bar s \gamma_\mu \gamma_5 s|N\rangle =
G_1^s (q^2) \bar {\cal U}_N \gamma_\mu\gamma_5 {\cal U}_N\, ,\quad\quad
 G_1^s (0) =\Delta s\, ,
\label{G1_s}
\end{equation}
where $G_1^s$ denotes the strange axial coupling of the nucleon.
\par
If chiral symmetry is spontaneously broken so that Eq.~(\ref{f_pi}) and 
its hypercharge analog are valid,
the nucleon matrix elements of the flavor preserving
isovector and hypercharge axial vector currents are
in turn dominated by pion-- and $\eta $ meson pole terms in the soft 
limit $k_\mu\rightarrow 0$. This means that the $\pi $ and $\eta $ meson 
are the respective isovector and octet Goldstone bosons whose
currents are partially conserved
\begin{equation}
\partial^\mu J^{(3)}_{\mu ,5} = (f_{\pi^0} m_{\pi^0})
m_{\pi^0}^2\phi^0_\pi \, ,\qquad
\partial^\mu J^{(8)}_{\mu ,5} =  (f_\eta m_\eta)
m_\eta^2\phi_\eta \, .
\label{PCAC_eta}
\end{equation}
These PCAC relations are clearly consistent with Eq.~(\ref{f_pi}) and 
its hypercharge analog. As a consequence,
the pseudoscalar coupling constants of the pion and the $\eta $ meson
to the nucleon (in turn denoted by $g_{\pi NN}$ and $g_{\eta NN}$)
satisfy Goldberger--Treiman type relations and are in the ratio
\begin{equation}
{g_{\eta NN}\over g_{\pi^0 NN}}
= { {g_A^{(8)}\, m_N }\over {(f_\eta m_\eta )}}:
  {{g_A\, m_N}\over {(f_{\pi^0} m_{\pi^0} )}}\,
\approx  {g_A^{(8)}\over g_A}\, .
\label{eta_const2}
\end{equation}
Here, the {\it ordinary} pion decay constant
($f_{\pi^0} m_{\pi^0} $) is taken to be approximately equal to the one
of the $\eta $ meson, $f_\eta m_\eta $.
{}From Eq.~(\ref{eta_const2}) it follows that the size of the
meson-nucleon coupling constant is another criterion to help one decide
whether or not a meson is to be considered as a Goldstone boson and what
kind of Goldstone boson it is.
\par
While for the pions experimental data are well known to have confirmed the
Goldberger-Treiman relation $f_\pi m_\pi g_{\pi NN}=g_A m_N$ to an accuracy
of about 6$\%$~\cite{CS}, this is not the case for $g_{\eta NN}$.
The SU(3) flavor symmetry and the NQM predict a relatively large value for
\begin{equation}
g_{\eta NN}={1\over \sqrt{ 3}}{{3\cal F -\cal D}\over {\cal F+\cal D}}
            g_{\pi^0 NN}={\sqrt{3}\over 5}g_{\pi^0 NN}\, ,
\label{eta}
\end{equation}
if the standard axial vector ${\cal F/\cal D}$ ratio $2/3$ is adopted.
The same value is obtained when the NQM spin fractions $\Delta u=4/3$,
$\Delta d=-1/3$, $\Delta s=0$ are used in
Eqs.~(\ref{gA}),(\ref{gA_0}), (\ref{eta_const2}).
On the other hand, the recent deep inelastic scattering experiment
E143\cite{E143} reports the following spin fractions for the proton
\begin{equation}
\Delta u=0.84\pm 0.05,\quad \Delta d=-0.43\pm 0.05,\quad
\Delta s=-0.08\pm 0.05,
\label{spinfr1}
\end{equation}
its analysis relying on the ${\cal F/\cal D}$ axial vector ratio of the 
SU(3) flavor symmetry. If these spin fractions are used in 
Eqs.~(\ref{gA}), (\ref{gA_0}), (\ref{eta_const2}) in conjunction 
with $g_A={\cal F+\cal D}=1.2573\pm 0.0028$,~\cite{PART} then
\begin{equation}
g_{\eta NN}=(0.26\pm 0.11)g_{\pi^0 NN}
\label{spfeta1}
\end{equation}
results, again a fairly large value. With the spin fractions
from~\cite{EK}
\begin{equation}
\Delta u=0.85\pm 0.03,\quad \Delta d=-0.41\pm 0.03,\quad
\Delta s=-0.08\pm 0.03,
\label{spinfr2}
\end{equation}
one finds
\begin{equation}
g_{\eta NN}=(0.28\pm 0.06)g_{\pi^0 NN},
\label{spfeta2}
\end{equation}
confirming the previous value with smaller error. Clearly, the more
negative $\Delta s$ is the larger $g_{\eta NN}$ will be if the
$\eta$ meson is taken to be the octet (or hypercharge) Goldstone boson.
\par
The difference between these estimates for the octet coupling 
$g_{\eta NN}$ is about 20 to 25$\%$. This is larger than
the pion deviation quoted earlier but of the same order as the deviation,
$\Delta _K =1-(m_N+m_{\Lambda})g_A^{(4\pm i5)}/(2f_K m_K g_{KN\Lambda})
\approx 30\pm15\%$, from the
Goldberger-Treiman relation for the kaons~\cite{CAD} and in line with the
mass scale $m_K\sim 3.5m_{\pi}$ to $m_{\eta }\sim 3.9m_{\pi}$. Using the
NQM octet values for $g_A^{(8)}$ and $g_{\eta NN}$ in the corresponding
deviation gives about 25$\%$ for the $\eta $ meson. Corrections from 
chiral perturbation theory are of order 30\%\cite{Savage} and therefore 
much too small to help one understand better the problem of the 
suppressed $\eta $NN coupling. 
\par
The size of the pseudoscalar $\eta $--nucleon coupling constant
$g_{\eta NN}$ has been repeatedly subjected to comparisons with data.
The analysis of  $\bar pp $ collisions by means of a dispersion relation
technique in \cite{GrKr} led to the surprising result of a vanishing
$\eta $-nucleon coupling. Later on, fitting nucleon--nucleon as well as
hyperon--nucleon phase shifts by means of the full Bonn potential, no
need for an $\eta $ meson exchange was found in \cite{Hol}, and
\cite{Reub}. Most recently, accurate differential cross sections
for $\eta $ photoproduction off protons at threshold obtained at
the Mainz Microtron (MAMI) were shown to require an $\eta $-nucleon
coupling constant suppressed by a factor of more than two relative to
the quark model value in Eq.~(\ref{eta}) giving~\footnote{
The small value for $g_{\eta NN}$ ensures cancellation between
the Born terms and the background processes proceeding via $\omega $
and $\rho $ intermediate states and leads to an almost 100\% dominance
of the $S_{11}$ resonance in case of S--wave $\eta $ photoproduction.
This allows one to determine the N(1535) parameters from
the $p(\gamma , \eta )p $ reaction at threshold. For details concerning 
the $g_{\eta NN}$ extraction procedure from data, the interested reader 
is referred to the original literature~\cite{Lothar}.}
\begin{equation}
|g_{\eta NN}| = 0.16 g_{\pi NN}\, .
\label{MAMI_eta}
\end{equation}
The violation of the octet Goldberger-Treiman relation is actually worse
than a naive comparison of Eqs.~(\ref{spfeta2}) and (\ref{MAMI_eta}) would
suggest because Eq.~(\ref{spfeta2}) has to be compared to the tree level
(contact) piece of $g_{\eta NN}$ in Eq.~(\ref{MAMI_eta}). We shall come
back to this point at the end of the next section.
\par
The comparison of the $\eta$-nucleon coupling constant predicted from the
octet Goldberger-Treiman relation with the data seems to rule out the 
$\eta$ meson as the octet Goldstone boson. Thus, one has to search for a
mechanism that supports the decoupling of the $\eta $ meson from the
nucleon. In the next section we demonstrate that the suppressed
$\eta N$ coupling constant can be understood if the $\eta $ decay
into the hadronic vacuum proceeds through the total neutral axial
vector current associated with the $Z$ boson rather than through
the hypercharge axial vector current of the SU(3)--flavor quark model.
Note that this total isoscalar axial vector current
can be represented as an ideal mixing between weak currents
having the same structure as the hypercharge and the
flavor singlet axial currents.
Within such a scheme the $\eta $- (and in addition the
$f_1$NN) vertex constants will appear proportional to $\Delta s$, the
fraction of nucleon spin carried by the strange quark sea, thus
explaining their suppression naturally.
A summary of the main results is provided in Sect. III.

\section{Decoupling of the $\eta $ meson from the octet axial vector current}

As a consequence of the Goldstone boson nature
of the $\eta $ meson its weak decay is usually supposed
to proceed via the hypercharge axial vector current. This means that one
ascribes to the $\eta $ meson the fundamental ability to filter out its
octet component from the isoscalar electroweak axial vector current, much
like an optically active material filters out a particular circularly
polarized component from linearly polarized light. We question this
viewpoint and argue that the experimentally
observed decoupling of the $\eta $ meson from the nucleon is naturally
understood if one assumes that the $\eta $ meson is coupled
to the hadronic vacuum as an ordinary neutral pseudoscalar
meson directly through the $Z$ boson.
The neutral gauge current (denoted by $J_\mu^{Z}$)
of the electroweak theory corresponding to $Z$ boson exchange
\cite{Georgi} is given by
\begin{eqnarray}
J_\mu^{Z} & =& -2 \sum_{i=1}^{3}
\bar\Psi _{iL} \gamma_\mu {\tau^{weak}_3\over 2} \Psi_{iL}
+2\sin^2 \theta_W J_{\mu }\, ,
\label{Z0_curr}\\
\Psi_{1L} =
\left(
\begin{array}{c}
u\\
d
\end{array}\right)_L\, , &&
\Psi_{2L} = \left(
\begin{array}{c}
c\\
s
\end{array}\right)_L\, ,
\qquad
\Psi_{3L} = \left(
\begin{array}{c}
t\\
b
\end{array}\right)_L\, .
\label{quark_gen}
\end{eqnarray}
Here $\Psi_{iL}$ denotes the $i$th lefthanded
quark generation, $\tau^{weak}_3/2$ stands for the weak isospin,
$\theta_W$ is the Weinberg angle,
and $J_\mu $ the electromagnetic current.
{}From Eq.~(\ref{quark_gen}) one immediately sees
that the first quark generation
acts simultaneously as a doublet both for weak and strong
isospin. This observation explains the success of
the current algebra statement on the identity between
the weak and strong isovector axial vector currents.
In contrast, the second and the third quark generations
decompose into singlets with respect to strong
isospin and so do their respective axial vector currents.
The neutral axial gauge current of the electroweak theory
(denoted by $J_{\mu ,5}^{GWS}$) that emerges from the first
term on the rhs in Eq.~(\ref{quark_gen}) reads
\begin{equation}
J_{\mu ,5}^{GWS} = -{1\over 2} \bar u \gamma_\mu \gamma_5
+{1\over 2} \bar d \gamma_\mu \gamma_5 d
+{1\over 2} \bar s \gamma_\mu \gamma_5 s +...
= -J_{\mu ,5}^{(3)} + J_{\mu ,5 }^{(s)} \, ,
\label{GWS_axial}
\end{equation}
where the contribution of heavier flavors was
ignored for simplicity.
Therefore, the weak isosinglet axial current is purely
strange. In the following we will adopt the viewpoint that
the hypercharge and the singlet axial currents are ideally mixed,
as are those of the electroweak theory
and work out the decay matrix element of the
$\eta $ meson (considered as the octet scalar state
$|\eta \rangle = {1\over \sqrt{6}} (\bar u u +\bar d d -2\bar s s)$
for simplicity).
Subsequently, we will make use of a scheme that was first exploited
by Jaffe in \cite{Jaffe} to calculate of the decay constant, $ f_M$,
of a vector meson (M).
The main assumption is that
the coupling of a component $(\bar q_i q_i)_M$ of a  $1^{--}$ vector
meson $M$ to a current $\bar q_j \gamma_\mu q_j$ is diagonal in flavor
\begin{equation}
\langle 0|{1\over 2} \bar q_i \gamma_\mu q_i|(\bar q_j q_j)_M\rangle
= \kappa_j (1^{--}) \delta _{ij} m^2_M\epsilon^M_\mu\, .
\label{Ja_cpl}
\end{equation}
In the spirit of flavor symmetry it is furthermore convenient to suggest
flavor independence of $\kappa _j$ at least for the light flavors,
i.e., $
\kappa_u(1^{--}) = \kappa_d (1^{--}) \equiv \kappa (1^{--})\, $.
The decay constants of the vector mesons
as calculated within Jaffe's scheme can be shown to equal to those
predicted by the NQM.

\par
A convenient extension of Jaffe's coupling scheme to the
case of pseudoscalar mesons is obtained by replacing one
mass power by the corresponding meson four-momentum.
In doing so, the pion
decay constant is found with the help of the relations
\begin{equation}
\langle 0|J^{GWS}_{\mu ,5}|\pi^0 \rangle
= f_{\pi^0} m_{\pi^0} ik_\mu \, ,
\qquad
f_{\pi^0}
= -\sqrt{2} \kappa (0^{--})\, ,
\label{eta_pi0}
\end{equation}
where $\kappa_u (0^{--}) = \kappa _d (0^{--}) \equiv \kappa (0^{--})$
was assumed, and
$|\pi^0 \rangle = {1\over \sqrt{2}} (\bar u u - \bar d d) $ used.
In a similar way the decay constant $f_\eta $ of the $\eta $
meson is calculated as
\begin{equation}
\langle 0|J^{GWS}_{\mu ,5}|\eta \rangle
= f_\eta m_\eta \, ik_\mu\, , \qquad
f_{\eta} = -{2\over \sqrt{6}}\,\kappa_s (0^{--}) \, .
\label{eta1_pi0}
\end{equation}
In case the flavor symmetry is violated, the values of $\kappa_s(0^{--})$
and $\kappa_{u/d}(0^{--})$ may differ appreciably. Indeed, from the
empirical values of $f_{\pi^0} m_{\pi^0} = 85\, $MeV and
$f_{\eta} m_{\eta} = 94\,$MeV \cite{PART} one
extracts $\kappa (0^{--}) = -0.44 $, and $\kappa_s(0^{--}) =-0.21 $,
respectively, which corresponds to the measured ratio of
\begin{equation}
{{f_\eta m_\eta} \over {f_{\pi^0} m_{\pi^0} } }\approx 1.1.
\label{fefp_ratio}
\end{equation}

Although the chiral perturbation theory calculation of Ref. \cite{Leutw}
predicts the value of $(f_\eta m_\eta)/(f_{\pi^0} m_{\pi^0} ) \approx 1.3$
in nice agreement with data, the analysis given above shows that the
closeness of the empirical $f_{\pi^0} m_{\pi^0} $ and $f_\eta m_\eta $
values is not necessarily indicative of SU(3) flavor symmetry but may just 
be brought about by the proper size of the coupling of the strange
quarkonium to the isosinglet axial vector current.
In case the hypercharge and the singlet currents are
ideally mixed, the isosinglet axial vector current of the nucleon
coincides with the weak isosinglet axial vector current
\cite{McKeo}\footnote{Note that in Ref. \cite{McKeo}
the amplitude of the isoscalar axial vector current
is defined as ${G_1^s\over 4}$ with $G_1^s = 2\Delta s$. }
\begin{equation}
\langle N|J_{\mu ,5}^{(s)}|N\rangle = {G_1^ s\over 2}
\bar {\cal U}_N \gamma_\mu \gamma_5 {\cal U} _N\, , \qquad G_1^s
= \Delta s.
\label{axcurr_nucl}
\end{equation}
The universal structure of the neutral axial vector current in both the
weak and strong interactions allows one to express the coupling constant
of the $\eta $ meson (at the tree level)
to any target by means of its weak decay constant as \cite{KiTi}
\begin{equation}
{g_{\eta NN}\over {m_N}} = { G_1^s\over {f_\eta m_\eta }}\, .
\label{etaNN_fin}
\end{equation}
Note that this equation is the 'strange' analog of the Goldberger-Treiman 
relation. Such expressions are commonly used in all (vector) meson dominance 
models and imply the standard assumption of unsubtracted dispersion relations 
for the form factors dominated by the corresponding poles.
Therefore, Goldberger--Treiman relations in fact reflect current universality.
For this deeper reason the quark model had suggested that the
$\eta $ meson weak decay proceed by the octet axial vector current, thus
keeping universality of the latter at both the strong and weak vertices.
{}With Eq.(~\ref{etaNN_fin}) the ratio $g_{\eta NN}/g_{\pi^0 NN}$ is
calculated as
\begin{equation}
|{g_{\eta NN}\over g_{\pi^0 NN}}|\approx
|{{\Delta s f_{\pi^0} m_{\pi^0}} \over {g_Af_\eta  m_\eta } }|
 \le  |{{\Delta s}\over {\Delta u -\Delta d}}|\sim  0.06\, ,
\label{etapi_ratio}
\end{equation}
where the values $\Delta s =-0.08$ and $g_A = 1.2573$ \cite{EK}
have been used.
Thus a natural explanation is found
for the suppression of the {\it contact} (tree level) $\eta $ nucleon
coupling.
\par
In view of the suppression of the tree level $\eta $-nucleon coupling,
loop corrections containing
non--strange mesons like the $a_0(980)\pi N$ triangular contribution
both to the $\eta NN $  and $f_1 NN$ vertices acquire
importance. Such corrections have first been considered in Refs.\cite{KiTi},
\cite{KiRi} and shown to be extremely useful to achieve agreement with
data on $\eta $ photo--production off protons at threshold.

\section{Conclusions}

\par
In the present study we reviewed the status of the
$\eta $ meson as the {\it octet} Goldstone boson of the spontaneously
broken $SU(3)_L\otimes SU(3)_R$ chiral symmetry. We showed that recent
precision experimental MAMI data on $\eta $ photoproduction off protons at
threshold fail to confirm the {\it octet} Goldberger--Treiman relation
for the pseudoscalar $\eta $-nucleon coupling constant.
The smallness of the experimentally observed $g_{\eta NN}$ compared to
its quark model value and another one implied by the proton spin fractions 
via $g_A^{(8)}$ was shown to have a natural explanation in assuming 
the $\eta $ meson to couple to a neutral axial vector current having
the same structure as the isosinglet axial current of the electroweak theory.
Within such a scenario the $\eta NN$ coupling constant follows from the
'strange' Goldberger-Treiman relation and, being proportional
to $\Delta s$, the fraction of nucleon spin carried by the strange
quark sea, the {\it contact}, or tree level, $\eta NN$ vertex appears
suppressed relative to its quark model value by the amount of
$\Delta s/ g_A^{(8)}$.

\par
The suppression of the nucleon matrix element of the hypercharge axial
current was considered in a series of papers with SU(3)
breaking \cite{Lipkin} on the one hand, and in the context of
both chiral perturbation theory \cite{Savage} and the
large $N_c$ limit of QCD \cite{Dai} on the other hand.
In particular,
in \cite{Dai} it was found that in the large $N_c$ limit
$3{\cal F} - {\cal D} = 0.28\pm 0.09\,$.
It is easy to verify that our idea of the decoupling of the
$\eta $ meson from the hypercharge axial vector current
does not contradict that small result provided it is properly
interpreted. Indeed, in our notation $3{\cal F} - {\cal D}$ equals
$\sqrt{3}\, g_A^{(8)}$. In case the $\eta $ meson couples to the purely
strange axial vector current only, $g_A^{(s)} :=\Delta s$ replaces
$g_A^{(8)}$ leading to $\sqrt{3}|\Delta s | \approx 0.14 $.
This is the value that has to be compared with $0.28\pm 0.09$ above.
It is important to remark that, if the reason for the
suppressed $g_{\eta NN}$ coupling is the preferred coupling of the 
$\eta $ meson to the strange axial vector current, then relating the
${\cal F}/{\cal D}$ ratio to the $\eta N$ system lacks any justification.
For this, the $\beta $ decay of the $\Xi^-$ hyperon is relevant. The
value~\cite{XS}
$3(G_A/G_V)_{\Xi^- \rightarrow \Lambda}=3{\cal F}-{\cal D}= 0.75\pm 0.15$
from data~\cite{PART} for this decay turns out to be much larger than the 
one associated with the experimentally determined $g_{\eta NN}$ coupling, 
so that from this angle as well the $\eta $ meson could not be the octet 
Goldstone boson.
\par
The difference between Eq.~(\ref{etaNN_fin})
and the experimental value of $g_{\eta NN}^2/4\pi  = 0.4 $ in
Eq.~(\ref{MAMI_eta}) can be
attributed to triangular $a_0\pi N$ vertex corrections along the line of
Ref.~\cite{KiTi}.
To recapitulate, our idea is to parametrize on the composite
hadron level the suppression of the $\eta NN$ vertex in terms of
the nucleon matrix element of the purely strange axial current,
$\langle N|J_{\mu, 5}^{(s)}|N\rangle $, that is subsequently
increased by effective triangular vertices containing
non--strange mesons thus accounting for the effective strangeness
of the surrounding meson cloud necessarily originating from  OZI
violation effects. This aspect appears to be more rigorous than
those considered in Refs.~\cite{Lipkin}, \cite{Savage} and \cite{Dai}
where the concept of the SU(3) flavor symmetry is still kept
despite substantial violation effects.
It is worthwhile to mention that, if the OZI rule applied to
the $0^{--}$ mesons (as it does for the $1^{--}$ mesons),
then the $\eta $ meson would be purely strange and its coupling to the
nucleon would be natural via the 'strange' Goldberger-Treiman relation 
given in Eq.~(\ref{etaNN_fin}). It is not so obvious to realize that,
although mass formulas deduced from the constituent quark model
prefer a small $\eta-\eta'$ mixing, the corresponding sea quark
currents could be ideally mixed and the 'strange' Goldberger-Treiman 
relation in Eq.~(\ref{etaNN_fin}) valid that gives the $\eta $ meson 
features of a 'strange' Goldstone boson.

\par
As a further consequence of the scheme considered, additional
information on the coupling constants of neutral axial vector mesons 
like the $f_1(1285)$ meson can be obtained: the $f_1 $ nucleon
coupling arises exclusively because of the violation of the OZI rule
within the axial vector meson nonet, and the $f_1$ mesons can not
be considered as the chiral partners of the $\omega $ and $ \phi $ 
mesons from the vector meson nonet. Thus, the $1^{--}$ and $1^{++}$ 
nonets can no longer be viewed as complete chiral partners.
\par
We also wish to stress that the decoupling of the $\eta $ meson
from the nucleon, as revealed by various experimental studies during the
last decade, questions the validity of SU(3)$_L\otimes $SU(3)$_R$ chiral
rotations as symmetry transformations of the quark flavor triplet.
Rather, two--flavor chiral rotations, SU(2)$_L\otimes $SU(2)$_R$, seem
to be valid that act on distinct I-, V- and U- spin quark doublets
and give rise to flavor changing currents. These strong currents
have the same structure (up to a Cabibbo rotation within the $U$
spin doublet) as the weak ones. Within this scenario the strong
and weak isosinglet axial vector currents have equal structures
too and don't contain any longer $u$ and $d$ quarks. Therefore,
the  $\eta  $ meson will couple to strange quarks only. Thus little room
is left for $\eta $ meson exchange among up and down quarks. Hence
isospin--independent spin--spin and tensor
forces between quarks should be attributed to instanton effects as
considered in \cite{HJW} rather than to $\eta $ meson exchange.
Finally, another point of this comment is
that probing the strange content of hadrons by strongly interacting
$\eta $ mesons might be as important as by weakly interacting
$Z$ bosons, possibly opening a field for new experimental activities.

\section{Acknowledgement}
It is a pleasure to thank Prof. Konrad Kleinknecht for
his interest and the critical reading of the manuscript.
The work of M.K. was supported by Deutsche Forschungsgemeinschaft
(SFB 201).

\end{document}